\documentstyle[epsfig,eqsecnum,aps]{revtex}
\begin{document}
\draft
\title{Gamow-Teller strength distributions in Xe isotopes}

\author{O. Moreno$\:^1$, R. \'Alvarez-Rodr\'{\i}guez$\:^1$, P. Sarriguren$\:^1$, 
E. Moya de Guerra$\:^{1,2}$, J. M. Ud\'{\i}as$\:^2$ and J. R. Vignote$\:^{1,2}$}
\address{$^1 \:$ Instituto de Estructura de la Materia.
Consejo Superior de Investigaciones Cient\'{\i }ficas. \\
Serrano 123, E-28006 Madrid, Spain}
\address{$^2 \:$ Dpto. F\'{\i }sica At\'{o}mica, Molecular y Nuclear. 
Facultad de Ciencias F\'{\i }sicas.\\
Universidad Complutense de Madrid. Avda. Complutense s/n, E-28040 Madrid, Spain.}

\date{\today}
\maketitle

\begin{abstract}

The energy distributions of the Gamow-Teller strength are studied for even-even
Xe isotopes with mass numbers from 124 to 142. A self-consistent microscopic
formalism is used to generate the single particle basis, using a deformed Skyrme
Hartree-Fock mean field with pairing correlations in BCS approximation. The
Gamow-Teller (GT) transitions are obtained within a quasiparticle random phase
approximation (QRPA) approach using a residual spin-isospin interaction in the
particle-hole and particle-particle  channels. We then discuss the pairing BCS
treatment and the determination of the $ph$ and $pp$ residual interaction
coupling constants. We study the GT$^+$ and GT$^-$ strength distributions for
the equilibrium nuclear shapes, which are an essential information for studies
of charge-exchange reactions and double-$\beta$ processes involving these
isotopes.\\

\end{abstract}


\section{Introduction}

The isotope $^{136}$Xe has been recently used as a moving target colliding with
a hydrogen gas jet in a first test to proof the feasibility of the EXL experimental
approach at FAIR-GSI \cite{EXL}. Charge-exchange (p,n) reactions on this Xe isotope
were consequently measured. Although this kind of facilities are intended to explore
highly unstable nuclei, some stable isotopes like those under study here are
normally used as initial test targets. Therefore, having reliable information on
their nuclear structure turns out to be of primary importance. From $^{136}$Xe, we
extend our study to other stable isotopes, $^{124-134}$Xe, and also to the
neutron-rich region $^{138,140,142}$Xe. The latter undergo $\beta^-$-decay and
their half-lives provide us with another piece of experimental information.

Gamow-Teller transition matrix elements can be extracted from the measured
forward-angle charge-exchange data \cite{xsect}. At high incident energies and
at forward angles, the nuclear states are probed at small momentum transfer.
Therefore, only the central parts of the isovector effective interaction contribute
to the cross-section. Furthermore, because of the small momentum transfer, a
multipole expansion leads to a simple relation between the measured 0$^o$ cross
sections and the corresponding allowed $\beta$-decay transition rates (L=0).
Assuming that even taking into account the projectile distortion effects in DWBA
the transition amplitudes would still approximately factorize into a nuclear
reaction part and a nuclear structure part, the 0$^o$ charge-exchange cross
sections are proportional to the corresponding Fermi and Gamow-Teller matrix
elements. The proportionality factor is parametrized in terms of a distortion
factor, a kinematic factor and a volume integral of the effective
interaction \cite{xsect}.

Theoretical GT$^{\pm}$ strength distributions, as the ones discussed here, can
therefore be used to predict cross sections of various charge-exchange reactions
under the appropriate kinematic conditions. Some of the charge-exchange reactions
corresponding to GT$^-$ processes are:\\
$^{136}$Xe(p,n)$^{136}$Cs ; $^{136}$Xe($^{3}$He,$^{3}$H)$^{136}$Cs ;
$^{136}$Xe(d,2n)$^{136}$Cs ;\\
while some examples of those corresponding to GT$^+$ transitions are:\\
$^{136}$Xe(n,p)$^{136}$I ; $^{136}$Xe($^{3}$H,$^{3}$He)$^{136}$I ;
$^{136}$Xe(d,2p)$^{136}$I. \\

In addition, the Xe isotopes are of special interest regarding double $\beta$
processes\cite{ejiri}. Different nuclear models have been developed by many
groups to calculate quantitatively the double $\beta$ matrix elements, as
described in recent review articles \cite{2b}. Table \ref{table1} shows all
the transitions of this type involving Xe isotopes as parents or daughters.
The GT$^{\pm}$ strength distributions which will be obtained here can be used
to calculate the transition amplitudes of the initial and final ground states
going to the virtual QRPA-excited states of the intermediate odd-odd nucleus.
After computing the overlap between the intermediate states coming from the
parent and from the daughter nuclei, the two-neutrino double beta decay matrix
elements as well as the half-lives can be calculated. Such a calculation has
indeed been carried out in \cite{alvarez} for $^{128}$Xe, $^{130}$Xe and
$^{136}$Xe, which are $\beta^- / \beta^-$ parents or daughters. 

Finally, Xe isotopes are of considerable interest because they belong to a typical
shape transitional region \cite{wyss}, in which there are experimental indications
of triaxial deformation in some isotopes \cite{lieb}. Present theoretical triaxial
calculations are mainly based on algebraic models \cite{pan}. Nevertheless, in this
work we assume axial symmetry. As we shall see, for the deformed Xe isotopes with
two equilibrium shapes there are no critical changes in the Gamow-Teller strengths
at these two shapes, and a similar behavior may be expected when considering
possible triaxial shapes. In any case, axial deformation is a crucial ingredient
of the formalism that gives rise to new features in the Gamow-Teller strength
distributions different from those obtained within a spherical treatment.

Although some of the isotopes studied may be spherical, in this work we
are not restricted to those but we deal with a large number of Xe
isotopes whose equilibrium shapes are unknown. For most of the isotopes
considered here, there is no clear experimental evidence of whether they
are spherical or deformed. Under these circumstances, a deformed approach
is always preferable over a spherical one because a deformed formalism
contains the spherical shape as a particular solution. 

We have also found in the past \cite{sarr,exp} that the GT strength distributions
may depend, in some cases significantly, on the deformation of the decaying
nucleus. But we notice that this dependence has to be studied case by case
since it is very sensitive to the fragmentation and crossing of levels
generated by the deformation. We think it is indeed worth studying the
degree of sensitivity of the GT strength of these Xe isotopes to deformation.

Following this introduction we present in Sec. II a brief description of the
theoretical framework. Sec. III includes our results regarding HF+BCS energies
and GT$^{\pm}$ strength distributions, together with a discussion on the pairing
treatment. Finally, Sec. IV contains the main conclusions of our work.

\section{Theoretical framework}

We describe here briefly the theoretical formalism used, whose details can be
found in ref. \cite{sarr}. We carry out a deformed Hartree-Fock calculation with
the effective nucleon-nucleon density-dependent Skyrme interaction Sk3 \cite{sk3},
assuming axial deformation and time reversal symmetry. The single-particle wave
functions are expanded in terms of the eigenstates of an axially symmetric
harmonic oscillator in cylindrical coordinates using eleven major shells.
Pairing correlations between like nucleons are included in BCS approximation
either taking fixed pairing gap parameters ($\Delta_{\pi}$ for protons and
$\Delta_{\nu}$ for neutrons) or taking fixed pairing strength parameters
($G_{\pi}$ and $G_{\nu}$ respectively). We refer to these two types of
calculations as HF(Sk3)+BCS($\Delta$) and HF(Sk3)+BCS(G), respectively.
They yield single particle energies and wave functions together with their
occupation probabilities for protons and neutrons separately. A quadrupole
constrained HF+BCS calculation \cite{constraint}, where the intrinsic
quadrupole moment is constrained, is also performed to obtain the deformation
dependence of the ground state energy.

The pairing energy gaps $\Delta$ are determined phenomenologically, and the
pairing strengths $G$ are obtained from them in an indirect way, as will be
described later on. Within BCS approximation, both parameters are related by
the so called gap equation:

\begin{equation}\label{gaps}
\Delta=G\sum_i u_iv_i \:,
\end{equation}
where $v_i$ y $u_i$ are occupation and non-occupation probability amplitudes of
the $i^{th}$ single particle level subject to the condition
$v_i^2\:+\:u_i^2\:=\:1$. In order to determine the value of $G$ that reproduces
a given value of $\Delta$, one should take into account that this depends on the
active energy range and number of levels considered. In our case, we include
all HF single particle sates in our basis above and below Fermi level. It is
important to stress that the occupation probability amplitudes are computed in
each iteration of the HF method, and are used to calculate the one-body density
and  mean field of the next iteration, so that one gets new single-particle
wave functions, energies and occupation numbers at each iteration. Therefore,
the selfconsistent determination of the binding energy and deformation includes
pairing correlations from the beginning.

To describe Gamow-Teller excitations we add to the quasiparticle mean field a
separable spin-isospin residual interaction in the particle-hole ($ph$) and
particle-particle ($pp$) channels, which is treated within QRPA. The advantage
of using separable forces is that the QRPA energy eigenvalue problem is reduced
to find the roots of an algebraic equation. The $ph$ part
\begin{equation}
V^{ph}_{GT} = 2\chi ^{ph}_{GT} \sum_{K=0,\pm 1} (-1)^K \beta ^+_K 
\beta ^-_{-K}, \qquad 
\beta ^+_K = \sum_{\pi\nu } \left\langle \nu \left| \sigma _K \right|
\pi \right\rangle a^+_\nu a_\pi \, ,
\end{equation}
is responsible for the position and structure of the GT resonance 
\cite{sarr,moller,hir,homma}. The corresponding coupling constant,
$\chi_{ph}^{GT}$ is obtained in a consistent way from the same energy density
functional as the Hartree-Fock mean field through a second derivative with
respect to the nucleonic density and averaging the contact interaction over
the nuclear volume \cite{sarr}. The $pp$ part consists of a proton-neutron
pairing force, which we introduce as a separable force \cite{hir,muto,engel},

\begin{equation}
V^{pp}_{GT} = -2\kappa ^{pp}_{GT} \sum_K (-1)^K P ^+_K P_{-K}, \qquad 
P ^+_K = \sum_{\pi\nu} \left\langle \pi \left| \left( \sigma_K\right)^+
\right|\nu \right\rangle  a^+_\nu a^+_{\bar{\pi}} \, .
\end{equation}
The coupling constant $\kappa_{pp}^{GT}$ may in principle be derived consistently
with the HFB or HF+BCS mean field through a second derivative with respect to the
pairing tensor of the energy density functional. This derivation would be
analogous to the way in which the ${ph}$ force is obtained as the second
derivative with respect to the density. In our theoretical scheme the
proton-neutron pairing interaction is neglected in the construction of the mean
field to avoid mixing of even-even and odd-odd isotopes in the intrinsic state.
Only pairing between like particles is included in the BCS approximation. This
implies that the particle-particle interaction in the proton-neutron channel is
in principle undetermined. Therefore the coupling constant $\kappa_{pp}^{GT}$
is fitted to the phenomenology, as for example to reproduce half-lives as it is
usually done \cite{hir,homma}.

The pnQRPA phonon operator for GT excitations in even-even nuclei is written as

\begin{equation}
\Gamma _{\omega _{K}}^{+}=\sum_{\pi\nu}\left[ X_{\pi\nu}^{\omega _{K}}
\alpha _{\nu}^{+}\alpha _{\bar{\pi}}^{+}-Y_{\pi\nu}^{\omega _{K}}
\alpha _{\bar{\nu}}\alpha_{\pi}\right]\, ,  
\label{phon}
\end{equation}
where $\pi$ and $\nu$ stand for proton and neutron, respectively,
$\alpha ^{+}\left( \alpha \right) $ are quasiparticle creation (annihilation)
operators, $\omega _{K}$ are the RPA excitation energies, and 
$X_{\pi\nu}^{\omega _{K}},Y_{\pi\nu}^{\omega _{K}}$ the forward and backward
amplitudes, respectively. It satisfies

\begin{equation}
\Gamma _{\omega _{K}} \left| 0\right\rangle=0\, ; \qquad
\Gamma ^+ _{\omega _{K}} \left| 0\right\rangle = \left|
\omega _K \right\rangle .
\end{equation}
when acting on the QRPA ground state of the parent nucleus
$\left| 0\right\rangle $.

The technical details to solve the QRPA equations are well described in
Refs. \cite{sarr,moller,muto}. Here we only mention that, because of the use
of separable residual forces, the solutions of the QRPA equations are found by
solving first a dispersion relation, which is an algebraic equation of fourth
order in the excitation energy $\omega$. Then, for each value of the energy,
the GT transition amplitudes in the intrinsic frame connecting the ground
state $\left| 0\right\rangle $ to one phonon states in the daughter nucleus
$\left| \omega _K \right\rangle $, are determined by using the normalization
conditions of the phonon amplitudes. They are given by

\begin{equation} 
\left\langle \omega _K | \beta _K^{\pm} | 0 \right\rangle = \mp 
M^{\omega _K}_\pm \, ,
\label{amplgt}
\end{equation}
where 
\begin{equation}
M_{-}^{\omega _{K}}=\sum_{\pi\nu}\left( q_{\pi\nu}X_{\pi\nu}^{\omega _{K}}+
\tilde{q}_{\pi\nu}Y_{\pi\nu}^{\omega _{K}}\right) \, ; \qquad 
M_{+}^{\omega _{K}}=\sum_{\pi\nu}\left( \tilde{q}_{\pi\nu}
X_{\pi\nu}^{\omega _{K}}+ q_{\pi\nu}Y_{\pi\nu}^{\omega _{K}}\right) \, ,
\end{equation}
with
\begin{equation}
\tilde{q}_{\pi\nu}=u_{\nu}v_{\pi}\Sigma _{K}^{\nu\pi };\ \ \ 
q_{\pi\nu}=v_{\nu}u_{\pi}\Sigma _{K}^{\nu\pi}\, ;\qquad
\Sigma _{K}^{\nu\pi}=\left\langle \nu\left| \sigma _{K}\right| 
\pi\right\rangle \, ,
\label{qs}
\end{equation}

It is a simple matter to find out that the Ikeda sum rule 
\begin{equation}
\sum_{\omega_K} \left[ \left( M_{-}^{\omega _{K}}\right) ^2- 
\left( M_{+}^{\omega _{K}}\right) ^2
\right] = 3(N-Z)
\label{ikeda}
\end{equation}
holds in RPA approximation provided all the eigenvalues contained in the
basis space are included in the sum, so that the orthonormalization conditions
are satisfied. In practice, the strength functions are calculated up to an
energy  $\omega < E_{\rm cut}$, where $E_{\rm cut}$ is such that Ikeda sum
rule is fulfilled up to a few per thousand discrepancy. Typical energies used
in our calculations are  $E_{\rm cut}=30$ MeV. The number of configurations
involved in this mass region for this energy range is typically over one
thousand.

Once the intrinsic amplitudes in Eq. (\ref{amplgt}) are calculated, the
Gamow-Teller strength $B(GT)$ in the laboratory frame for a transition 
$I_i K_i (0^+0)\rightarrow I_fK_f(1^+K)$ can be obtained as

\begin{equation}
B^{\pm}(GT)= \sum_{M_i,M_f,\mu} \left| \left< I_fM_f \left| \beta ^\pm _\mu
\right| I_i M_i \right> \right|^2= \left\{ \delta_{K_f,0} \left< \phi_{K_f} 
\left|  \beta ^\pm _0 \right| \phi_0\right> ^2 +2\delta_{K_f,1} \left< 
\phi_{K_f} \left|  \beta ^\pm _1 \right| \phi_0\right> ^2 \right\} \, .
\label{streven}
\end{equation}
To obtain this expression we have used the initial and final states in the
laboratory frame expressed in terms of the intrinsic states $|\phi_K >$,
using the Bohr and Mottelson factorization \cite{bm}.
Theoretical $\beta^-$-decay half-lives are calculated by summing up all the
energetically allowed transition probabilities in (\ref{streven}), in units
of $g_A^2/4\pi$, weighted with phase space factors, up to states in the
daughter nucleus with excitation energies below the $Q_{\beta^-}$ window.

One may wonder whether, in deformed nuclei, the calculated GT strengths
may contain spurious contributions from higher angular momentum components
in the initial and final wave functions. As mentioned above, the GT strengths
are calculated in the laboratory frame in the factorization approximation
of Bohr and Mottelson. Using angular momentum projection techniques
\cite{moya}, the angular momentum projection can be carried out through an
expansion in inverse powers of the angular momentum operator component
perpendicular to the symmetry axis $<J_{\perp}^2>$. This expansion, to lower
order, provides a factorization approximation formally identical to that of
Bohr and Mottelson. Thus, the effect of angular momentum projection is to a
large extent taken into account. An upper bound to contributions from higher
angular momentum components is proportional to $<J_{\perp}^2>^{-2}$, with
values of $<J_{\perp}^2>$ ranging from 10 to 40 in the case of the deformed
Xe isotopes. Therefore, exact angular momentum projection in deformed Xe
isotopes would lead in all cases to less than a few percent effect in the
GT strengths. In the cases where the shape is spherical, there are not high
angular momentum contributions to the GT strengths.

It may be questioned whether it is correct to introduce additional BCS
parameters and residual interactions on top of the Skyrme interaction,
which is already an effective interaction. Indeed, if the effective force
used in constructing the mean field were to be the most general possible
interaction, one should not include additional parameters. However, this 
is not the case for the Skyrme interaction. The parameters of the Skyrme
force are determined by requiring that they reproduce the nuclear
compressibility, as well as the total binding energies and charge radii of
magic nuclei in spherical selfconsistent calculations. It is well known that
the effective Skyrme interaction and its existent parametrizations are suitable
to generate the optimal HF mean field of spherical and deformed nuclei
\cite{sk3,skyrmes}. The particle and hole eigenstates of the mean field, which
form the canonical basis, have highly nontrivial wave functions that contain a
mixture of many harmonic oscillator shells, when expanded into a harmonic
oscillator basis. In addition to deformation, for non-closed shell nuclei,
one has to take into account the pairing correlation effects which are
important when the level density around the Fermi level becomes large.

It is also well known that the effective Skyrme interaction is not suitable
to generate the quasiparticle mean field, since in the fit of the Skyrme
forces no attention is paid to the realistic character of the pairing
matrix elements, and extensions of the Skyrme HF method have been developed
over the years. To include pairing correlations in the mean field one
possibility is to do Hartree-Fock-Bogoliubov calculations using either
finite range forces, like the Gogny force, or contact density dependent
pairing interactions \cite{pairing}. All of them are extensions of the
Skyrme forces specifically designed for this purpose. The other possibility
is to perform BCS calculations in the canonical basis using either
phenomenological fixed gap parameters or fixed pairing strengths, as
originally proposed by Vautherin. This is the path followed in our paper.
This path has been proved to be successful to study the properties of ground
state and low spin excited states in open shell nuclei
\cite{sarr,sk3,skyrmes,m1}.

Concerning the residual interactions, we mentioned above that the same
effective Skyrme interaction is used to generate a separable particle-hole
residual interaction. The separable interaction simplifies enormously the
calculation and still contains the main characteristics of the contact
force. The quasiparticle energy density functional obtained with the effective
Skyrme and pairing interactions that we used does not contain any
dependence on particle-particle interactions in the proton-neutron
channel. Therefore we have to introduce a proton-neutron particle-particle
residual interaction in the usual way as a separable force with a 
coupling strength that we fit to the measured half lives. We would like to
recall that a bridge between Skyrme HF and RPA calculations for excited
states was established long time ago \cite{tsai} by using a particle-hole
force in the RPA,  which is determined by the second derivatives of the HF
energy with respect to the density. But there is no guarantee that the derived
force is good for excited states. This is so because Skyrme forces are
constructed for the description of ground state properties. Finally, we would
like to notice that similar schemes based on Skyrme HF+BCS+RPA have been
frequently used in the literature with successful results \cite{hama}.

\section{Results and discussion}

\subsection{Equilibrium deformations}

A deformed Hartree-Fock mean field calculation is performed using a Sk3 Skyrme
force with a constraint of the quadrupole deformation given by the parameter
\mbox{$\beta\:=\:\sqrt{\pi/5} \: Q_p/(Z\langle r^2\rangle)$}, where $Q_p$ is
the proton quadrupole moment and $\langle r^2\rangle$ is the charge mean square
radius. A pairing interaction between like nucleons within BCS approximation is
also included, keeping fixed the pairing energy gap ($\Delta$) or the pairing
strength ($G$).  From this calculation we obtain the ground state energy as a
function of the quadrupole deformation $\beta$.

In principle, when a pairing force with fixed strength $G_{\pi,\nu}$ is used,
the pairing energy gaps $\Delta_{\pi,\nu}$ depend on the strength of the
pairing interaction as well as on the occupation amplitudes of the
single-particle states (see Eq.(\ref{gaps})). This last condition amounts to
say that pairing energy gaps are deformation-dependent. Actually, we obtain
these gaps first phenomenologically from the odd-even mass differences by
means of a symmetric five term formula involving experimental binding energies
\cite{audi}, and keep them fixed to carry out a deformation-constrained
HF+BCS($\Delta$) calculation. Next we fix the pairing strength so as to
reproduce the gap parameters at the deformation of the ground state, and
we carry out a deformation-constrained HF+BCS(G) calculation. In this way
we perform a fixed pairing strength calculation, which is conceptually more
appealing, but still profiting from the experimental information available
for the gaps.

Fig. \ref{HF+BCS} shows the HF+BCS energies for $^{124-142}$Xe, using Sk3 Skyrme
interaction. The pairing interaction is included in both fixed gap (dashed line)
and fixed strength (solid line) treatments, each of them in a different curve
whose absolute minima have been separated 1 MeV for a better comparison. For
the same isotope, both curves show energy minima at very similar deformations,
the ones corresponding to the fixed pairing strength treatment being slightly
smaller in absolute value. One can also observe that the energy barrier at
the spherical region of $^{124-132}$Xe and of $^{140-142}$Xe is less pronounced
when the pairing strength is fixed, and in this case both minima have a very
similar energy (the prolate one being generally the ground state, except for
$^{126}$Xe). As the mass number increases from A=124 to A=138, the deformations
of the equilibrium shapes decrease and eventually converge to a spherical shape.
The two final isotopes show only one equilibrium deformation in the prolate
region.

Underneath each of these energy-deformation graphs we plot the corresponding
pairing gaps as a function of the quadrupole deformation for the fixed pairing
strength calculation. The vertical lines join the ground states from fixed gap
treatment with the corresponding pairing gaps at this deformation from the
fixed strength treatment. These values are those coming from the aforementioned
odd-even experimental mass differences.

In our calculations with fixed $G$ values the binding energies show a tendency
to increase at those deformations where the pairing gaps reach a minimum. This
looks contradictory since one may expect the opposite because the smaller the
pairing gap is, the lower is the contribution of the pairing energy to the
total binding energy. However, it indicates that the minima of the pairing
gaps appear at similar deformation to that where the volume and the spin-orbit
term contributions to the binding energy are maximum. On the other hand,
comparing calculations with G fixed to those with $\Delta$ fixed, one sees
in \ref{HF+BCS} that for $\beta$ values where $\Delta$ takes larger values
the binding energy increases more compared to that obtained with lower
($\Delta$ fixed) value.

In Table \ref{table2} we show the quadrupole deformation $\beta$ of the equilibrium
shapes according to the HF(Sk3)+BCS($\Delta$) and HF(Sk3)+BCS(G) calculations.
We also show for comparison the results from independent theoretical calculations
obtained from selfconsistent relativistic calculations \cite{ring} as well as
from phenomenological nonrelativistic calculations \cite{moller95}. These results
also indicate the existence of deformed solutions, which agree with those obtained
here. Upper limits of the ground state deformation obtained from experimental
B(E, 2$^+_1 \to$ 0$^+_1$) transitions are also included \cite{raman}.
Table \ref{table3} shows the pairing gaps, obtained from experimental binding
energies, and the pairing strengths reproducing these gaps at the ground state
deformation.

\subsection{Gamow-Teller strength distributions}

The spin-isospin residual interactions in the particle-hole and particle-particle
channels are treated here within a quasiparticle random phase approximation. The
particle-hole residual interaction coupling constant $\chi_{ph}$ is obtained
consistently with the Hartree-Fock mean field, and their values vary from 0.21 MeV
in $^{124}$Xe to 0.19 MeV in $^{142}$Xe. We have used an average value for all the
isotopes under study, $\chi_{ph}$ = 0.2 MeV. In the case of the particle-particle
residual interaction, we have chosen the value of the coupling constant
$\kappa_{pp}$ so as to reproduce the experimental half-lives of the three
unstable Xe isotopes: $^{138}$Xe (T$_{1/2}$=844.8 s), $^{140}$Xe
(T$_{1/2}$=13.60 s) and $^{142}$Xe (T$_{1/2}$=1.22 s). A good agreement between
calculated and experimental $\beta^-$ half-lives for the three isotopes is reached
with $\kappa_{pp}$=0.07 MeV, provided we use the standard attenuation factor
0.77 for spin matrix elements as in previous works.

The single particle energies and occupation probabilities at the equilibrium nuclear
shapes are obtained from a HF(Sk3)+BCS(G) calculation. Fig. \ref{GTminus} shows
GT$^-$ strengths (in $g_A^2/(4\pi)$ units) as a function of the excitation energy
of the daughter nucleus after the transition. Discrete and gaussian-folded
distributions are shown, the latter being more suited to compare with experimental
results regarding the GT strengths themselves or the cross sections of
charge-exchange reactions obtained from them. The range of the  excitation energy
from 0 to 30 MeV includes the resonance, which appears at around 13 MeV in
$^{124}$Xe and moves slightly toward higher excitation energies as the mass
number increases, reaching 25 MeV in $^{142}$Xe.

Fig. \ref{GTplus} shows the same calculations but for GT$^+$ transitions. As
expected from the Ikeda sum rule, Eq. (\ref{ikeda}), the scale of the strengths is
much smaller in this case. Table \ref{table4} contains the summed
GT$^{\pm}$ strengths, their difference, the value of $3(N-Z)$ and the fulfillment
of the Ikeda sum rule in percentage for the prolate shape of every isotope
(which is generally the ground state), and for the oblate shape when there is
a second minimum. The Ikeda sum rule is fulfilled up to a very high degree of
accuracy in all the cases.

The fragmentation observed in the GT$^+$ strength is reduced as the prolate energy
minimum moves to the spherical region ($^{134,136,138}$Xe). A double peak structure
becomes then apparent, which was responsible for the larger width of these
resonances in comparison with the ones in GT$^-$ distributions. A second peak
with less strength, in the near spherical shapes, appears 8 MeV below the biggest
one and moves accordingly with it as the mass number changes. It is worth noticing
the large single peak which appears at very low excitation energies in $^{126}$Xe
and also in $^{124}$Xe, reaching in the prolate shape of this last case a strength
of 0.34 $g_A^2/(4\pi)$, as indicated in the figure. This strength corresponds to
a dominant GT transition from a $K^\pi = 9/2^+$ proton state to a $K^\pi = 7/2^+$
neutron state connecting the proton $g_{9/2}$ shell with the neutron $g_{7/2}$
shell. The occupation probability of the neutron state is small enough to allow
the transition in $^{124}$Xe and $^{126}$Xe, but when the number of neutrons
increases, this state becomes blocked for GT transitions. 

From the GT$^{\pm}$ profiles of $^{124-132}$Xe it is obvious that no clear
distinction can be made between oblate and prolate deformations. It is only
possible to distinguish both deformations in some cases when small energy windows
are explored, as for example in the low B(GT$^+$) energy window of $^{124,126}$Xe.
Similar studies on the effect of deformation in the GT strength distributions were
done in the neutron-deficient Hg-Pb-Po region \cite{more}, and in the A$\simeq$70
mass region \cite{sarr}.

The effect of deformation on the GT strength distributions can be observed 
more clearly in Fig. \ref{new}, where we compare spherical and deformed QRPA
results. We show the examples of $^{128}$Xe, where two equilibrium shapes, oblate
and prolate, are obtained, and $^{140}$Xe, where a prolate shape is predicted.
In Fig. \ref{new} plots upward correspond to
deformed calculations with prolate shapes, while plots downward correspond to
spherical calculations. The left panels show the GT$^-$ strengths and the right
ones the GT$^+$ strengths. As we can see in Fig. \ref{new} the main effect of
deformation is the stronger fragmentation of the strength, which is particularly
clear on the GT$^+$ strength distributions because of the smaller scale. The
positions of the peaks are also changed from spherical to deformed in a different
way for each case.

In order to compare the GT strength distributions obtained here with those coming
from a HF(Sk3)+BCS($\Delta$) calculation of the single particle levels, we show
in Fig. \ref{comp1} the distributions corresponding to both BCS pairing treatments
(fixed gap and fixed strength) for the prolate and oblate equilibrium shapes of
$^{128}$Xe. The results are very similar, as could be expected given the fact
that the equilibrium nuclear shapes occur at very similar deformations and binding
energies in both BCS treatments (see Fig. \ref{HF+BCS}). In particular, for the
prolate shape both distributions are almost identical as expected because the
values of the $G_{\pi,\nu}$ parameters were chosen to reproduce the
$\Delta_{\pi,\nu}$ parameters precisely at this deformation. It is interesting
to know that also at the oblate minimum these pairing strengths nearly
reproduce the values of the phenomenological pairing gaps, since as seen in
Fig. \ref{HF+BCS} they are very close to the ones at the prolate minimum.
Therefore the distributions for the oblate case with both pairing treatments
are also very similar. In the case of spherical equilibrium shapes, as for
example $^{136}$Xe, the GT strength distributions from fixed pairing gap and
fixed pairing strength calculations are indistinguishable.

\section{Summary and conclusions}

The Xe isotopes are of considerable theoretical interest because they participate
in a variety of double-beta decay processes and because they belong to a nuclear
shape transitional region. In addition, from the experimental point of view, the
stable Xe isotopes have been used as moving targets in charge-exchange reactions
to test new facilities, where the unstable Xe isotopes will be explored in the
near future. The present work has addressed these topics by predicting stable
nuclear shapes and GT strength distributions, which are a fundamental tool to
calculate single and double beta transition matrix elements and half-lives,
as well as cross sections of charge-exchange reactions under the appropriate
kinematic conditions. 

For even-even Xe isotopes with mass numbers from 124 to 142, we have studied
the GT strength distributions using a deformed pnQRPA formalism with $ph$ and
$pp$ spin-isospin separable residual interactions. The quasiparticle mean field
is obtained from an axially deformed HF approach, with the Skyrme interaction
Sk3, including pairing correlations between like-nucleons in BCS approximation
using either fixed gaps ($\Delta_{\pi,\nu}$) or fixed pairing interaction
strength ($G_{\pi,\nu}$). The HF+BCS mean field has been also used to consistently
determine the $ph$ coupling constant for every isotope, whose average value has
been finally used for all of them. The $pp$ coupling constant has been fixed to
approximately reproduce the half-life of the three $\beta^-$-unstable Xe isotopes
included in this work.

From the energy-deformation curves, an oblate-prolate shape coexistence is
predicted in $^{124-132}$Xe with a low energy barrier between them. $^{134-138}$Xe
are predicted to be spherical, whereas the prolate shape seems to be strongly
favored in $^{140-142}$Xe. In general, a fixed pairing strength calculation
increases the binding energy of the spherical shape region. The deformation
dependence of the pairing energy gaps from a fixed pairing strength calculation
has been also shown. The self-consistent quadrupole deformations of the ground
state derived within the HF+BCS procedure are in agreement with independent
theoretical calculations as well as with the experimental upper bounds extracted
from B(E2) transitions in the whole chain of Xe isotopes under study. 

The GT$^-$ strength distributions for the equilibrium shapes are dominated by a
single peak moving to higher excitation energies and gathering more strength as
the number of neutrons increases, as expected. In the cases where two equilibrium
shapes are predicted, there is no strong dependence of the GT strength distribution
on the equilibrium shape, at least when a wide range of excitation energy is
considered. 

In the case of the GT$^+$ transitions, the strength is more fragmented, giving rise
to a richer structure in the energy distribution. This fragmentation decreases as
the energy minima move to the spherical region, where a double-peaked resonance
appears as observed in $^{134-138}$Xe. The lightest isotopes, $^{124}$Xe and
$^{126}$Xe, show a very high peak from a single transition at very low excitation
energies, but this transition is blocked in the isotopes with higher number of
neutrons. With such a complex structure, the influence of the sign of the nuclear
deformation (oblate or prolate) in $^{124-132}$Xe is more apparent on the
GT$^+$ strength distributions than on the GT$^-$ transitions, but in any case it
does not seem to be critical. However, the GT strength distributions obtained
from spherical or deformed shapes show different features related to the 
energy location of the main peaks and to the fragmentation of the strength.
From the accumulated GT$^{\pm}$
strengths up to 30 MeV of excitation energy, it has been shown that the Ikeda
sum rule is fulfilled up to a very high percentage (see table \ref{table4}). 

Theoretical studies of GT strengths and related observables as, in particular,
charge-exchange reaction cross sections are necessary to help in the event
simulation work of these kind of processes, that will be measured at the new
FAIR-GSI facility. Theoretical work on this direction is in progress.

\begin{center}
{\Large \bf Acknowledgments} 
\end{center}
This work was supported by Ministerio de Educaci\'on y Ciencia (Spain) under
contracts FIS2005-00640 and BFM2003-04147-C02-01. O.M. thanks Ministerio de
Educaci\'on y Ciencia (Spain) for financial support. R.A.R. thanks I3P Programme
(CSIC, Spain) for financial support. We also acknowledge participation in the
European Collaborations EURONS (RII3-506065), ILIAS (RII3-CT-2004-506222), and
INTAS-03-54-6545.

\newpage

\begin{center}

\begin{table}[t]
\caption{Double beta processes involving Xe isotopes, with their experimental
half-lives \protect\cite{tret,bara} and $Q$-values from experimental masses
\protect\cite{audi}. When a $\beta^+ / \beta^+$ transition is indicated,
$\beta^+ / EC$ and $EC / EC$ are also allowed (with $Q$-values 1.022 MeV
and 2.044 MeV lower than the one shown, respectively).}
\label{table1}
\begin{tabular}{rccc}
 Transition & 2$\beta$-process  & $T_{1/2}$ (yr) exp.  &$Q$-value [MeV]  \cr
\hline
$\mathbf{^{124}}$\textbf{Xe} $\to$ $^{124}$Te & $\beta^+ / \beta^+$  &
$>2.0\times 10^{14}$ \cite{tret} &2.866   \cr
$\mathbf{^{126}}$\textbf{Xe} $\to$ $^{126}$Te & $EC / EC$  & --- & 0.897  \cr
$^{128}$Te $\to$ $\mathbf{^{128}}$\textbf{Xe} & $\beta^- / \beta^-$  & 
$=2.5\pm0.3\times 10^{24}$\cite{bara} & 0.867   \cr
$^{130}$Te $\to$ $\mathbf{^{130}}$\textbf{Xe} & $\beta^- / \beta^-$  & 
$=0.9\pm0.1\times 10^{21}$\cite{bara}&2.529  \cr
$^{130}$Ba $\to$ $\mathbf{^{130}}$\textbf{Xe} & $\beta^+ / \beta^+$  & 
$>4.0\times 10^{21}$\cite{tret}&2.610   \cr
$^{132}$Ba $\to$ $\mathbf{^{132}}$\textbf{Xe} & $EC / EC$  &
$>3.0\times 10^{20}$\cite{tret} &0.840    \cr
$\mathbf{^{134}}$\textbf{Xe} $\to$ $^{134}$Ba & $\beta^- / \beta^-$  & 
$>1.1\times 10^{16}$\cite{tret} & 0.830  \cr
$\mathbf{^{136}}$\textbf{Xe} $\to$ $^{136}$Ba & $\beta^- / \beta^-$  &
$8.1\times 10^{20}$\cite{tret}  &2.468    \cr
\end{tabular}
\end{table}


\begin{table}[t]
\caption{Quadrupole deformation $\beta$ of the $^{124-142}$Xe equilibrium shapes
from a HF(Sk3)+BCS calculation obtained with fixed pairing gaps $\Delta$ as well
as with fixed pairing strengths $G$. Results from Refs.\protect\cite{ring} and 
\protect\cite{moller95} are also given for comparison.
Also given are the experimental values
obtained from B(E2) transitions \protect\cite{raman}.}

\label{table2}
\begin{tabular}{rcccccccc}
 A  & \multicolumn{2}{c} {$\beta_{th.}$ ($\Delta$ fixed)} & \multicolumn{2}{c} 
{$\beta_{th.}$ ($G$ fixed)} & $\beta_{th.}$ \cite{ring} & $\beta_{th.}$ 
\cite{moller95}& $|\beta_{exp.}|$ \cr
\cline{2-3} \cline {4-5}
& prolate & oblate & prolate & oblate &&& 
\footnotesize{from B(E, 2$^+_1 \to$ 0$^+_1$)}  \cite{raman}  \cr
\hline
124 & 0.24 & -0.19 & 0.22 & -0.17 & 0.215 & 0.208 & 0.264 (8) \cr
126 & 0.19 & -0.18 & 0.18 & -0.15 & 0.186 & 0.170 & 0.1881 (30) \cr
128 & 0.16 & -0.16 & 0.15 & -0.12 & 0.160 & 0.143 & 0.1837 (49) \cr
130 & 0.13 & -0.13 & 0.11 & -0.10 & 0.128 & -0.113 & 0.169 (6)   \cr
132 & 0.11 & -0.10 & 0.07 & -0.10 & -0.070 & 0.000 & 0.1409 (46) \cr
134 & 0.05 & -0.05 & 0.01 & -     & 0.000 & 0.000 & 0.120 (10)  \cr
136 & 0.00 & -     & 0.00 & -     & -0.001 & 0.000 & 0.086 (19)  \cr
138 & 0.03 & -     & 0.01 & -     & -0.002 & 0.000 & 0.0309 (18) \cr
140 & 0.15 & -0.10 & 0.12 & -     & 0.104 & 0.116 & 0.1136 (25)  \cr
142 & 0.17 & -0.12 & 0.16 & -     & 0.141 & 0.145 & -            \cr 
\end{tabular}
\end{table}


\begin{table}[t]
\caption{Pairing parameters for the HF+BCS calculation in $^{124-142}$Xe. Pairing
gaps $\Delta_{\nu, \pi}$ are obtained from experimental binding energies
\protect\cite{audi}. The pairing strengths $G_{\nu , \pi}$ are those reproducing
these gaps at the ground state deformation, which is also indicated (see text
for details).}
\label{table3}
\begin{tabular}{rcccccc}
 A  & $\Delta_{\nu}$ [MeV] & $\Delta_{\pi}$ [MeV] & $\beta_{th.}$ gs & 
$G_{\nu}$ [MeV] & $G_{\pi}$ [MeV] \cr
\hline
124 & 1.32 & 1.35 & 0.24 & 0.114 & 0.132 \cr
126 & 1.31 & 1.33 & 0.19 & 0.111 & 0.133 \cr
128 & 1.27 & 1.32 & 0.16 & 0.108 & 0.130 \cr
130 & 1.25 & 1.31 & 0.13 & 0.108 & 0.129 \cr
132 & 1.18 & 1.24 & 0.11 & 0.107 & 0.125 \cr
134 & 1.01 & 1.12 & 0.05 & 0.107 & 0.119 \cr
136 & 1.44 & 0.98 & 0.00 & 0.121 & 0.112 \cr
138 & 1.00 & 1.20 & 0.03 & 0.098 & 0.120 \cr
140 & 0.96 & 1.06 & 0.15 & 0.086 & 0.144 \cr
142 & 1.03 & 1.06 & 0.17 & 0.084 & 0.116 \cr
\end{tabular}
\end{table}


\begin{table}[t]
\caption{Calculated summed GT$^-$ and GT$^+$ strengths (in $g_A^2/(4\pi)$ units)
for the ground state and the first 0$^+$ excited state (in brackets) of
$^{124-142}$Xe, from a HF(Sk3)+BCS($G$) calculation. The \mbox{difference}
between both summed strengths is compared with the value of $3(N-Z)$ to check
the fulfillment of the Ikeda sum rule (in percentage).}
\label{table4}
\begin{tabular}{rcccccc}
 A  & $\Sigma GT^{-}$ & $\Sigma GT^{+}$ & $\Sigma GT^{-}-\Sigma GT^{+}$ & 
$3(N-Z)$ & $\%$ \cr
\hline
124 & 48.62 & 1.15 & 47.47 & 48 & 98.90 \cr
    & (48.34) & (0.84) & (47.50) & & (98.96) \cr
126 & 54.33 & 0.78 & 53.55 & 54 & 99.17 \cr
    & (54.23) & (0.71) & (53.52) & & (99.11) \cr
128 & 60.16 & 0.60 & 59.56 & 60 & 99.26 \cr
    & (60.18) & (0.61) & (59.57) & & (99.27) \cr
130 & 66.12 & 0.54 & 65.58 & 66 & 99.36 \cr
    & (66.15) & (0.56) & (65.59) & & (99.38) \cr
132 & 72.12 & 0.51 & 71.61 & 72 & 99.46 \cr
    & (71.83) & (0.46) & (71.37) & & (99.13) \cr
134 & 78.14 & 0.48 & 77.65 & 78 & 99.56 \cr
136 & 83.59 & 0.47 & 83.12 & 84 & 98.95 \cr
138 & 88.92 & 0.44 & 88.48 & 90 & 98.31 \cr
140 & 94.69 & 0.65 & 94.04 & 96 & 97.96 \cr
142 & 100.46 & 0.43 & 100.03 & 102 & 98.07 \cr
\end{tabular}
\end{table}

\newpage

\begin{figure}[t]
\epsfig{file=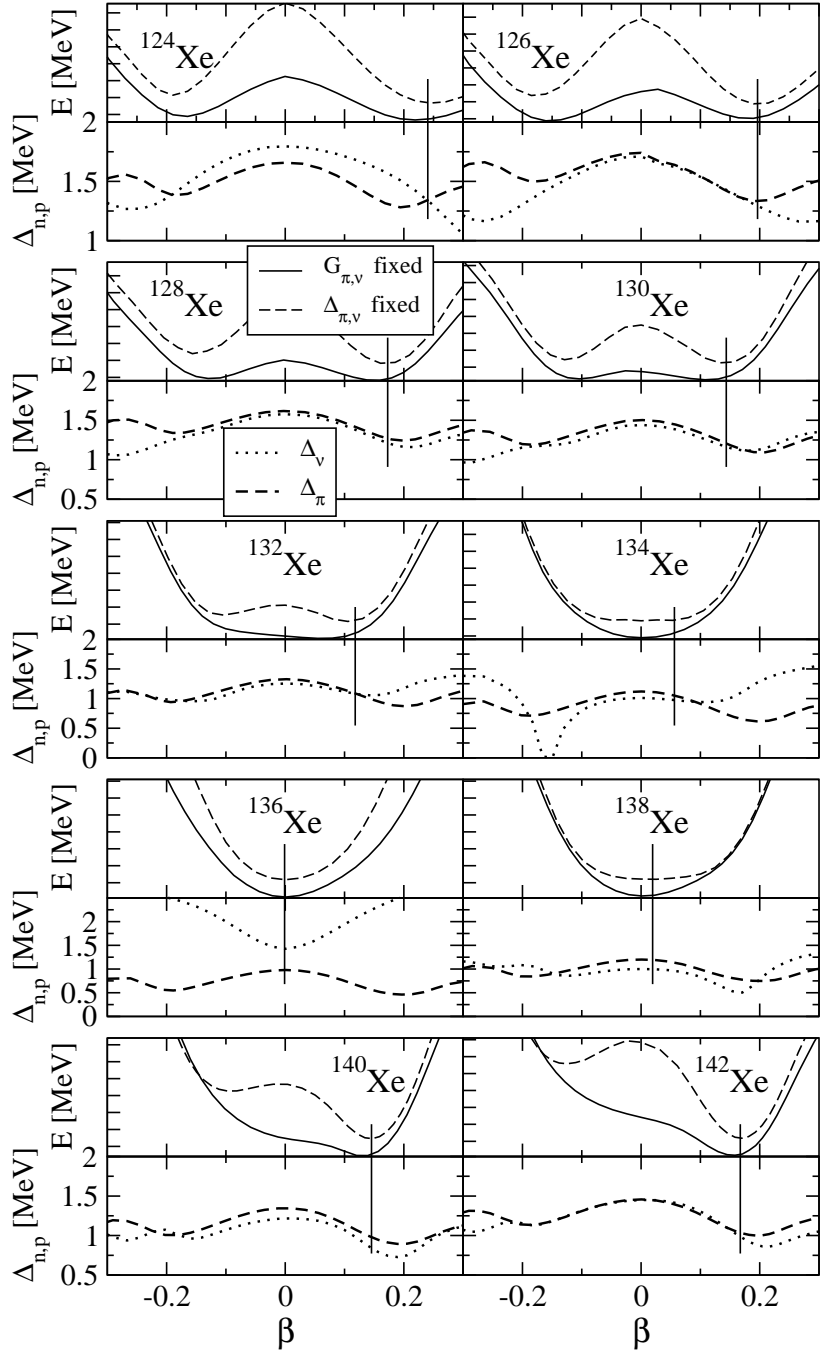,width=0.8\textwidth}
\vskip 1cm
\caption{HF(Sk3)+BCS energy of the isotopes $^{124-142}$Xe as a function of the
quadrupole deformation $\beta$ for fixed pairing gap (dashed line) and for fixed
pairing strength (solid line) treatments (with 1 MeV of separation between
absolute minima), as well as deformation dependence of pairing gaps from the
fixed strength calculation (dashed line for proton gap, dotted line for neutron
gap). The scale in the vertical axis is 1 MeV between two ticks. Vertical lines
indicate ground states from the fixed gap calculation.}
\label{HF+BCS}
\end{figure}

\newpage

\begin{figure}[t]
\epsfig{file=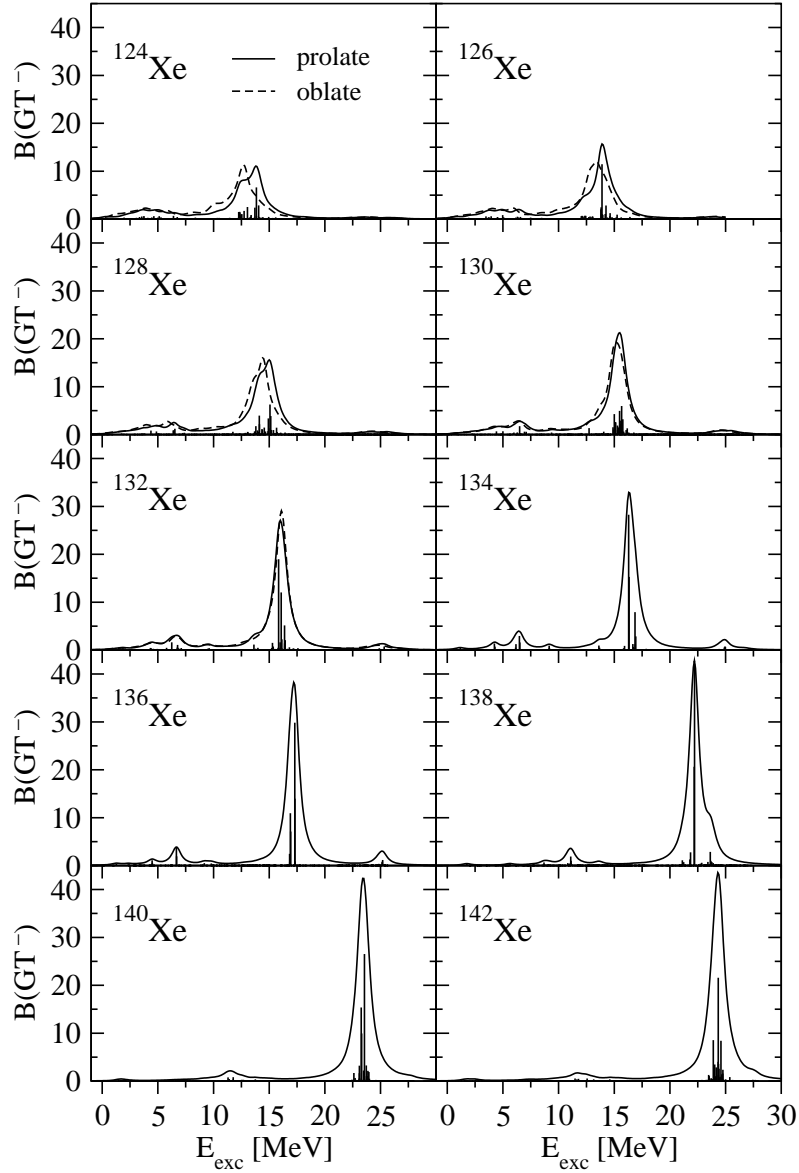,width=0.8\textwidth}
\vskip 1cm
\caption{Discrete and gaussian-folded Gamow-Teller strength distributions B(GT$^-$)
for $^{124-142}$Xe, from a HF(Sk3)+BCS(G) calculation. Two equilibrium deformations
appear for $^{124-132}$Xe, the solid line and the discrete spectrum corresponding
to the prolate shape, and the dashed line corresponding to the oblate shape.
The calculations for $^{134,136,138}$Xe correspond to spherical shapes, while those
for $^{140,142}$Xe correspond to prolate shapes (see Table \ref{table2}).}
\label{GTminus}
\end{figure}

\newpage

\begin{figure}[t]
\epsfig{file=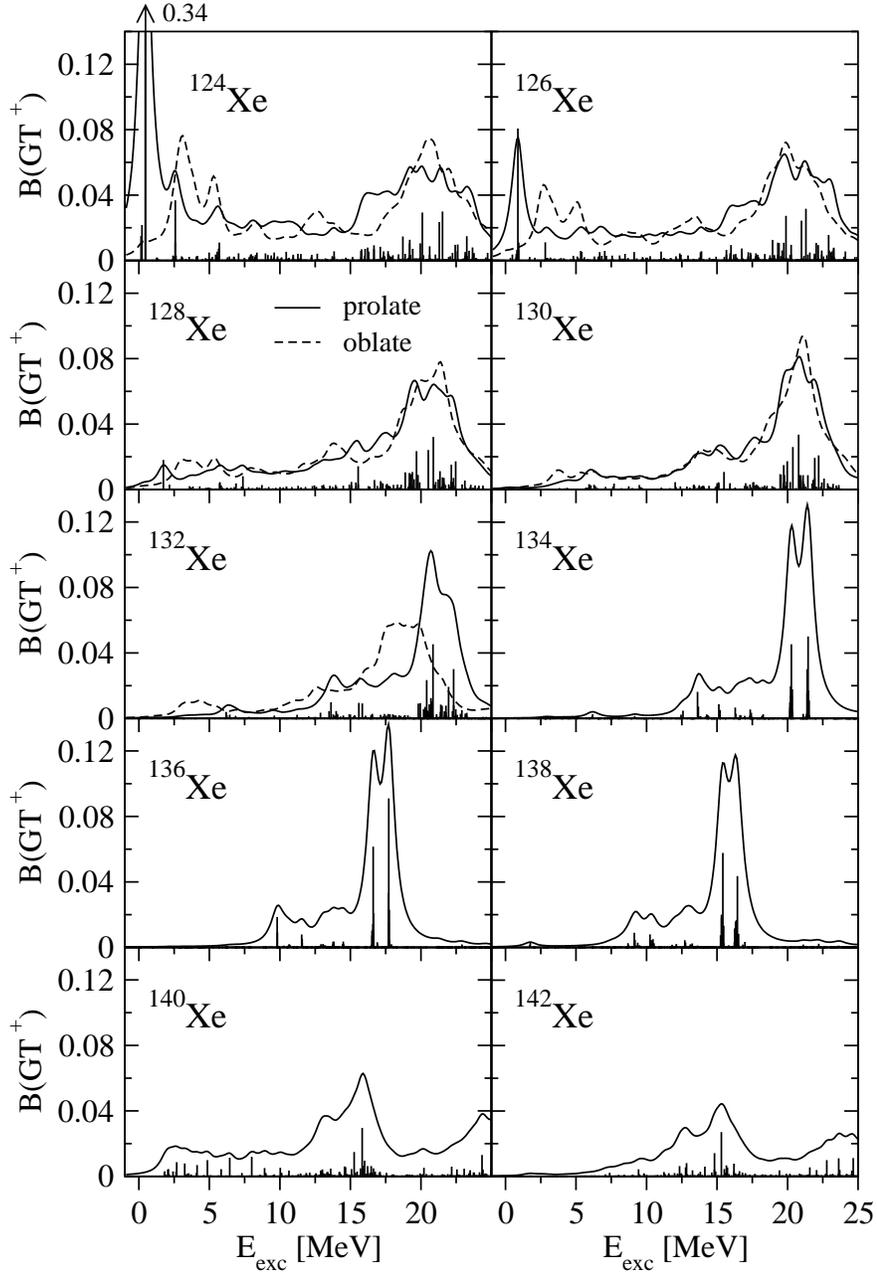,width=0.85\textwidth}
\vskip 1cm
\caption{The same as in fig. \protect\ref{GTminus} but for GT$^+$ strength
distributions.}
\label{GTplus}
\end{figure}

\newpage

\begin{figure}[t]
\epsfig{file=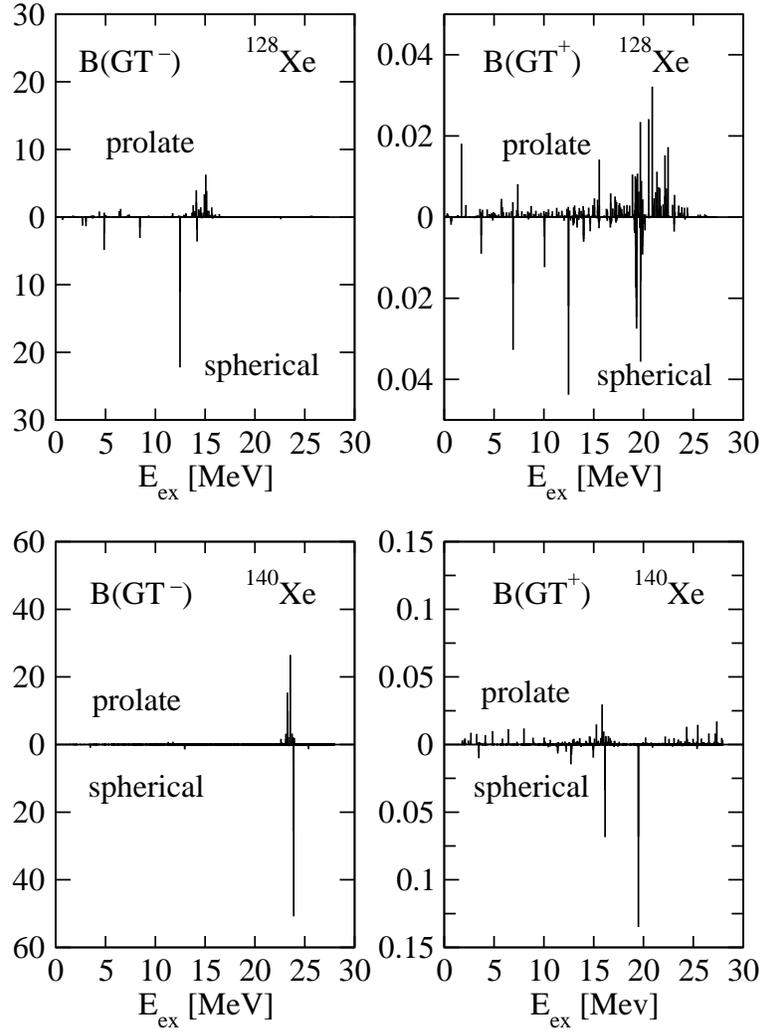,width=0.85\textwidth}
\vskip 1cm
\caption{GT strength distributions in $^{128}$Xe (upper panels) and 
$^{140}$Xe (lower panels). Left panels show GT$^-$ strengths, while
right panels show GT$^+$ strengths. Calculations with prolate (spherical)
shapes are plotted upward (downward), respectively.}
\label{new}
\end{figure}

\newpage

\begin{figure}[t]
\epsfig{file=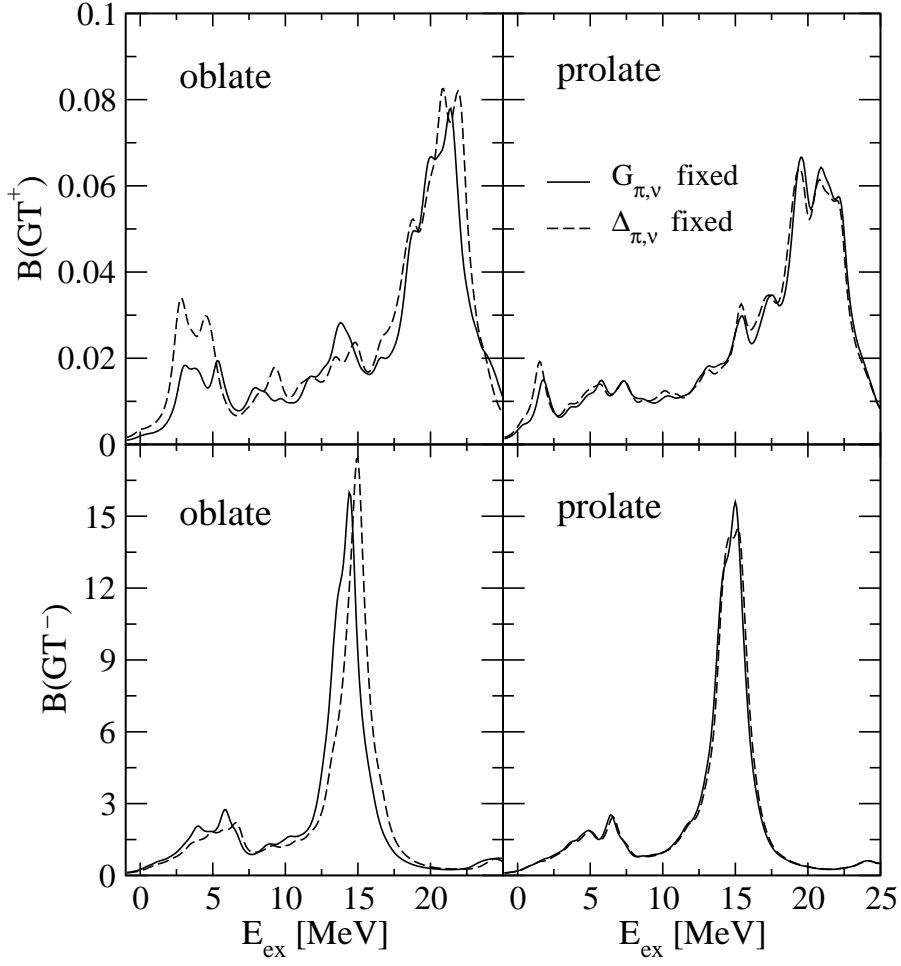,width=0.8\textwidth}
\vskip 1cm
\caption{Gaussian-folded Gamow-Teller strength distributions B(GT$^\pm$) in
$g_A^2/(4\pi)$ units for both equilibrium shapes of $^{128}$Xe, from a HF(Sk3)+BCS(G)
calculation (solid line) and a HF(Sk3)+BCS($\Delta$) calculation (dashed line).}
\label{comp1}
\end{figure}

\end{center}

\end{document}